\RequirePackage{ifpdf}
\pdfoutput=1 
\documentclass{JINST}

\usepackage{siunitx}
\usepackage{pdflscape}
\usepackage{gensymb}
\usepackage{graphicx}
\usepackage{pifont}
\usepackage{color}

\title{Electron drift in a large scale solid xenon}
\author{J.~Yoo\thanks{Corresponding Author: yoo@fnal.gov} ~ and W.~F.~Jaskierny\\
Fermi National Accelerator Laboratory, Kirk and Pine St., Batavia, IL 60510, USA}

\abstract{A study of charge drift in a large scale optically transparent solid xenon is reported. A pulsed high power xenon light source is used to liberate electrons from a photocathode. The drift speeds of the electrons are measured using a 8.7\,cm long electrode in both the liquid and solid phase of xenon. In the liquid phase (163\,K), the drift speed is 0.193 $\pm$ 0.003 cm/$\mu$s while the drift speed in the solid phase (157\,K) is 0.397 $\pm$ 0.006 cm/$\mu$s at 900 V/cm over 8.0\,cm of uniform electric fields. Therefore, it is demonstrated that a factor two faster electron drift speed in solid phase xenon compared to that in liquid in a large scale solid xenon.
}
\maketitle

\begin{document}
\section{Introduction}\label{intro}

\par Noble elements in both the gas and liquid phases have proven to be excellent low background radiation detectors~\cite{aprile2010,ackerman2011,aprile2012,akerib2014,kamland2013,PhysRevLett.113.121301}. In particular, xenon has drawn special attention among the noble elements due to several distinct advantages. The liquid phase of xenon possesses a very high scintillation light yield (40$\sim$60\, photons/keV) and the vacuum ultraviolet (VUV) wavelength (178\,nm) of the scintillation is optically transparent in xenon~\cite{jortner1965,Szydagis:2011tk}. The absence of long-lived radioisotopes in xenon results in no intrinsic background radiation sources. The large atomic mass of xenon (A=131.3 and Z=54) results in excellent self-shielding effects of radioactive backgrounds. The chemical purification of xenon is straightforward with the use of hot getter or gas distillation systems. A wide range of applications has been studied on particle tracking and spectroscopy including $\gamma$-ray astronomy, neutrinoless double beta decay, dark matter searches and neutrino coherent scattering experiments.

\par The solid phase of xenon has several advantages over liquid xenon. The density of solid xenon (3.41 g/cc) is higher than that of liquid xenon (2.95 g/cc), while the solid phase is transparent to its scintillation lights ($\sim$172\,nm)~\cite{jortner1965, Baum:1988, Varding1994, Kubota1982}. The electron drift speeds in solid xenon are measured to be faster compared to those in the liquid phase~\cite{Miller:1968zza}, which can be understood as suppressed electron-phonon scattering in the solid phase due to the reduced energetic phonon populations in low temperature media. Therefore, one may imagine a compact, scintillating, and fast ionization detector using the solid xenon. 

\par Particle detector applications based on the solid phase of noble element have been investigated extensively ~\cite{Miller:1968zza, Gushchin1982, Aprile1994129, Bolozdynya1977, Himi1982, Kubota1982, Kink1987, Bald1962, Varding1994, Michniak2002387, Aprile:1985xz}. Even though these studies have successfully demonstrated that the solid noble elements might be excellent candidates for a particle detector, large scale detectors have yet to be realized. In a previous publication, we demonstrated a scalability of above a kilogram scale of optically transparent solid xenon~\cite{yoo:2015sya}. The electro-negative contaminants, microscopic defects, voids and structural deformation are the dominant components that affect charge transportation. Therefore, it is non-trivial to maintain the fast charge transport properties in a large scale solid. 

\par The goal of this study is to measure the drift speed of electrons in a large scale optically transparent solid xenon. The sections below describe the experiment setup, results of electron drift velocity measurements, discussions about the results, and conclusion of the study.

\section{Experiment setup}\label{sec:experiment}

\begin{figure}[t!]
\begin{center}
\includegraphics[width=3.8in]{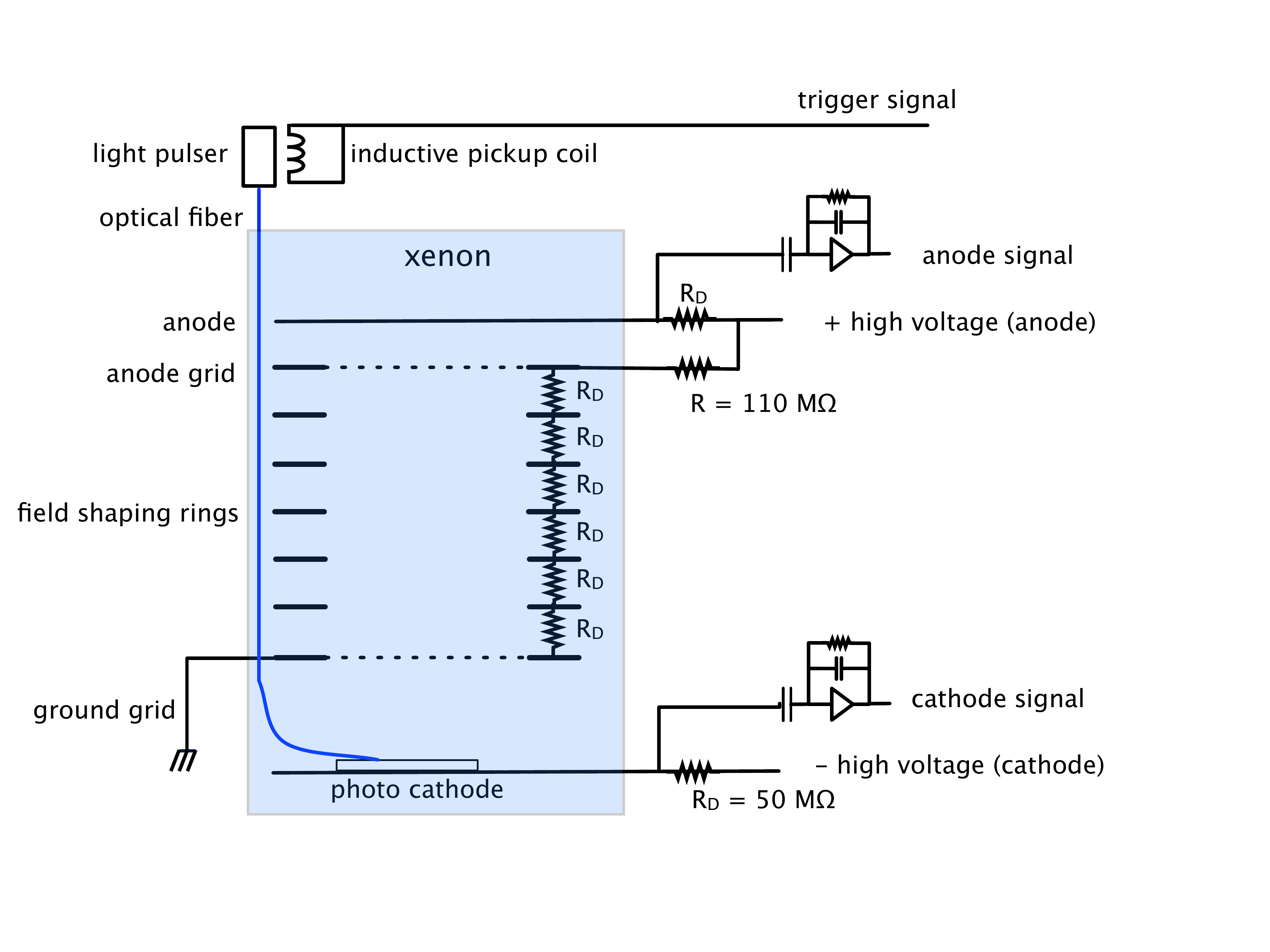}
\includegraphics[width=2.0in]{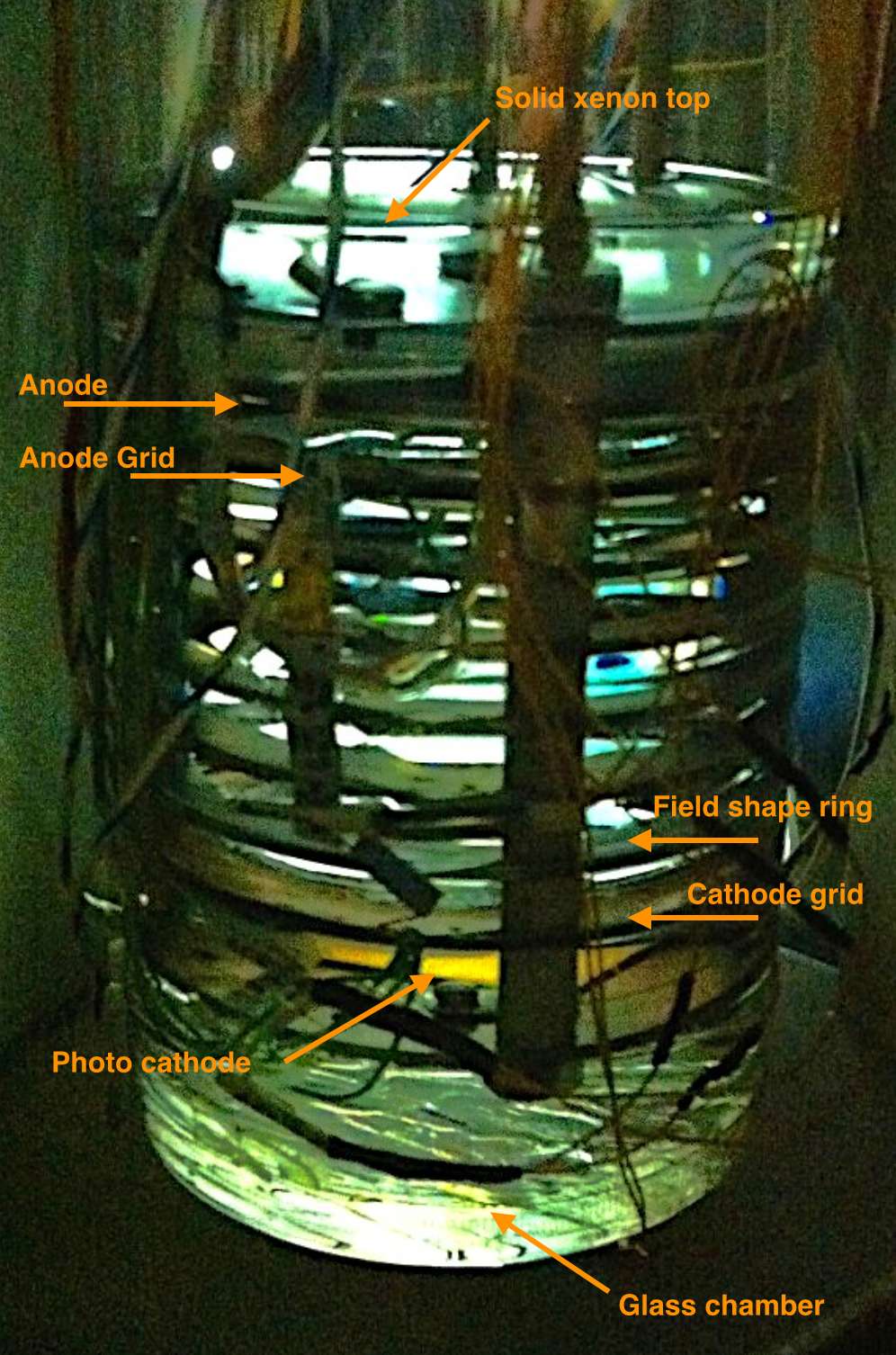}
\caption{A schematic diagram of the charge drift electrode (left) and a photograph of the electrode in solid xenon (right). The photons emitted from the optical fiber to the photo-cathode can be viewed from the external window. In this particular sample, there are small opaque spots near the edges of the field shape rings. Birefringence is visible near the bottom edge of the glass wall. However, the central area of the electrode where the electrons are drifting is optically transparent.~\label{fig:tpc}}
\end{center}
\end{figure}

\par The solid xenon test setup is designed to maximize the visual access of the cryogenic xenon volume. The cryostat consists of a stainless steel vacuum jacketed chamber with an outermost diameter of 30\,cm with three 15\,cm diameter glass window ports. Two concentrically placed glass chambers reside inside allowing optical access to the xenon bulk volume. The larger of the two glass chambers, which is used as a liquid nitrogen bath for cooling and has a diameter of 23\,cm, is referred to as the liquid nitrogen chamber. The smaller 10\,cm diameter inner chamber houses the xenon volume and is made out of Pyrex with a 5\,mm thick side wall and a 10\,mm thick flat bottom, is referred to as the xenon chamber. A commercial hot getter (PF4-C3-R-1 and Monotorr PS4-MT3-R-1 by SAES) and a circulation loop allows continuous purification of the xenon. The temperature and pressure are controlled using a programmable logic control system. 

\par The cryogenic and gas control of the system largely overlap with those of liquid-based systems, but additional fine tuning is required to solidify the xenon. Due to the density change between the liquid and the solid phase, the freezing process requires extra effort in order to not damage the detector components. A modified {\it Bridgeman's technique}~\cite{RGSv2} was adopted to grow solid xenon above a kg-scale. Details of the production of optically transparent solid xenon using this setup can be found in reference~\cite{yoo:2015sya}. Here we briefly explain the solid xenon growing process. First, xenon gas is condensed to liquid in the chamber at a set temperature of 163\,K and pressure of 14.5$\pm$0.5\,PSIA. When the liquid xenon level has reached above the top electrode plate, the bottom of the chamber is slowly cooled down to 145$\pm$0.5\,K while the barrel part of the chamber is cooled to 157$\pm$0.5\,K. Then the pressure of the xenon chamber is set to 17\,PISA. The xenon slowly solidifies from the bottom of the glass chamber. It takes more than two days to solidify the 2.3\,kg of xenon while keeping the optical transparency. After the solidification, the bottom temperature is slowly raised to the set temperature (157\,K) while keeping the barrel temperature at 157\,K. 

\par The charge drift electrode design is adopted from reference~\cite{Carugno1990580}. The design also employed a purity monitor that was developed for the liquid argon time projection chamber at Fermilab~\cite{Adamowski:2014daa}. It consists of parallel, circular electrodes, a disk supporting a photocathode, two open wire grids, one anode and one cathode, and anode disks for the field shaping. The grid support rings are made of G-10 material, and the grids are made of electro-formed, gold-sheathed tungsten with a 2.0\,mm wire spacing and 25\,$\mu$m wire diameter. The geometrical transparency of the grid is 98.8\%. The anode grid and the field-shaping rings are connected to the cathode grid by an internal chain of 50\,M$\Omega$ resistors. The photocathode which is attached at the cathode disk is an aluminum plate coated with 50$\si{\angstrom}$ of titanium and 1000$\si{\angstrom}$ of gold. A powerful (5 Joule) xenon flash lamp with a wide ultra-violet spectrum was used as the light source. The pulsed light is guided into the xenon chamber using two 0.6\,mm diameter core optical fibers. 
The edge of the fibers are directed to the center of the photocathode using a guide structure made by G-10 material. The distance between the edge of fibers to the photocathode is set at 1\,mm. Owing to the geometrical spread, the effect of the reflected photons producing photoelectrons is estimated to be less than 0.1\% of the primary photoelectrons. Therefore the contribution of those reflected photons to the charge drift time measurement is negligible. FIG.~\ref{fig:tpc} (left) shows a schematic diagram and a photograph of the charge drift electrode. The distances between the photocathode surface and the cathode-grid is 1.8\,cm, cathode-grid and anode-grid is 6.2\,cm with five field shape rings, anode-grid and anode is 0.7\,cm. Therefore, the full drift distance of electrons from the photocathode to the anode is 8.7\,cm. FIG.~\ref{fig:tpc} (right) shows a photograph of the electrode in solid xenon. 

\par A trigger signal is produced using an inductive pickup coil on the power leads of the flash lamp. The photo-electrons that are liberated from the photocathode drift to the cathode grid (1.8\,cm). After crossing the cathode grid, the electrons drift between the cathode grid and the anode grid (6.2\,cm). An electric current is induced at the anode after the electrons pass through the anode grid. The signals are read out using charge amplifiers which have a 5\,pF integration capacitor and a 22\,M$\Omega$ resistor in parallel. The capacitor provides a typical time constant of 110\,$\mu$s. A Tektronix 3034C Digital Phosphor Oscilloscope is used to readout the cathode and anode traces. For each trace, a total of 10$^4$ samples are read out over a 400\,$\mu$s of time window. The oscilloscope is connected to a PC through a GPIB communication protocol. A LabVIEW software is used to record the traces in a text file.

\section{Charge transport}\label{sec:chargetransport}
\par The pulsed light from the high power xenon lamp propagates through a 5\,m long optical fiber. The propagation time of the light in the optical fiber is about 25\,ns. The power and the frequency of the xenon arc lamp light source are set to 4 Joule per pulse and 10\,Hz, respectively. The light emittance to the photocathode is visually observable through the glass window. High voltages on each electrode are applied to achieve uniform electric fields over the cathode to anode. A negative high voltage is applied between the cathode and the cathode-grid (ground), and positive high voltages are applied from the cathode-grid to the anode-grid and from the anode-grid to the anode. The electrons are drifted from the cathode to the anode. The electric current is read out at the cathode disk as the {\it cathode signal}. No current can be measured between the cathode grid and the anode grid. The current between anode grid to anode is then read out as the {\it anode signal}. Therefore, the time difference between the trigger signal and the start time of the anode signal is the measurement of the electron drift time from the photo-cathode to the anode grid (8.0\,cm){\footnote{Due to the HV glitch near t=40$\mu$s, fitting results of cathode signals show large uncertainties in the $t_0$ measurement. Since the photon delivery time from the light source to the photocathode is 25ns, we conclude that the best $t_0$ values can be obtained by the timing of the HV glitch.} 

\begin{figure}[t!]
\begin{center}
\includegraphics[width=4.8in]{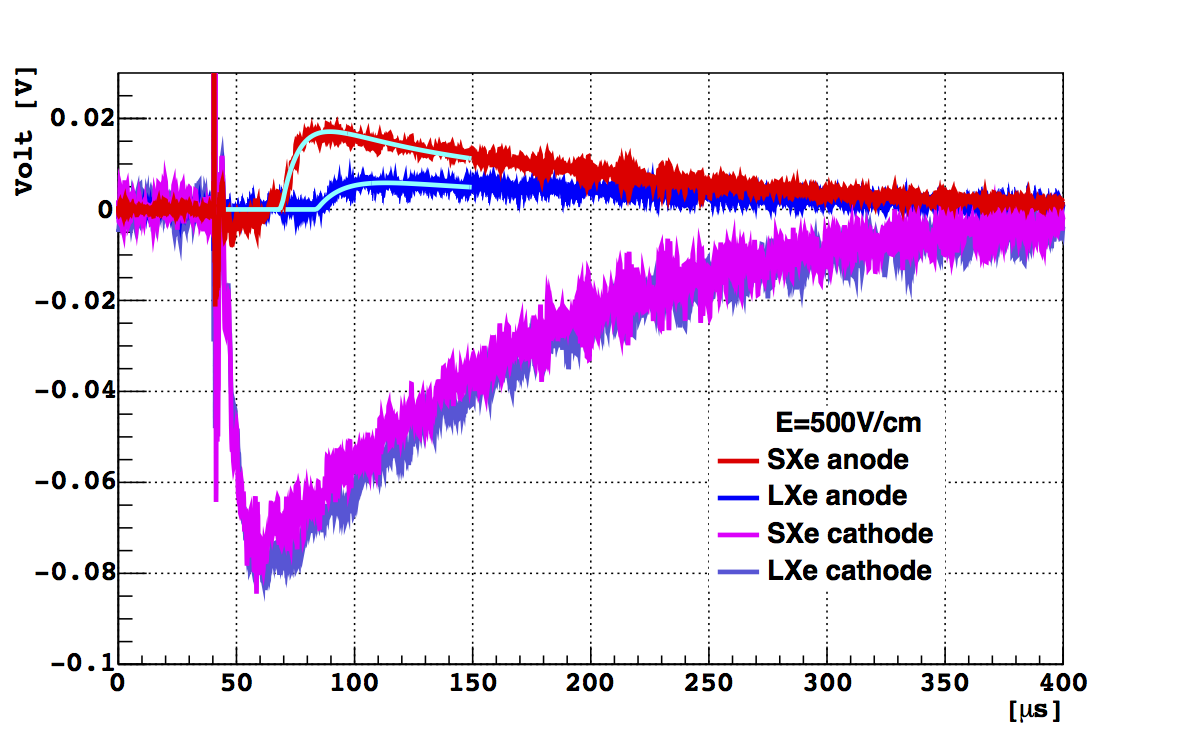}
\caption{Examples of traces in the liquid (163\,K, blue trace) and the solid (157\,K, red trace) phase of xenon. The applied electric field in these examples is 500\,V/cm. The glitch peaks at 40\,$\pm 0.5 \mu$s are induced by the high voltage trigger of the pulsed light source unit. The anode signal in the solid xenon is faster than that in the liquid xenon. The amplitudes of the cathode signals are similar in liquid and solid xenon. The functional form fit results are overlaid on the traces in light-blue curves. The charge drift distance between the photocathode (trigger signal) to the start time of the anode signal is 8.0\,cm.~\label{fig:tracecomp}}
\end{center}
\end{figure}

\par The drift velocity is strongly temperature dependent within ten degrees around the melting point. The electro-negative impurities in solid xenon has not been measured and can be different from the value of the liquid phase. Thus a direct measurement of the impurity concentration of the solid xenon was not possible. In order to avoid this difficulty, the electron drift time at 500 V/cm electric field is initially measured in liquid xenon. The xenon gas is continuously circulated through the getter system above the liquid xenon volume. The measurements were repeated until the anode signal amplitude did not improve over time, which normally takes about two weeks. Once the anode amplitude has stabilized (i.e. purity level of xenon is saturated), the electron drift time is measured in various electric fields (200\,V/cm to 900\,V/cm). After completion of the measurement, solid xenon was grown from the liquid phase of xenon. The solidifying process takes about a week including two-days of net solid xenon growing time. Once the entire xenon volume has solidified, the electron drift times in various electric fields are measured in the solid phase. The measurements are repeated after melting the solid xenon in order to confirm the original measurements in the liquid phase. 

\begin{figure}[t!]
\begin{center}
\includegraphics[width=2.9in]{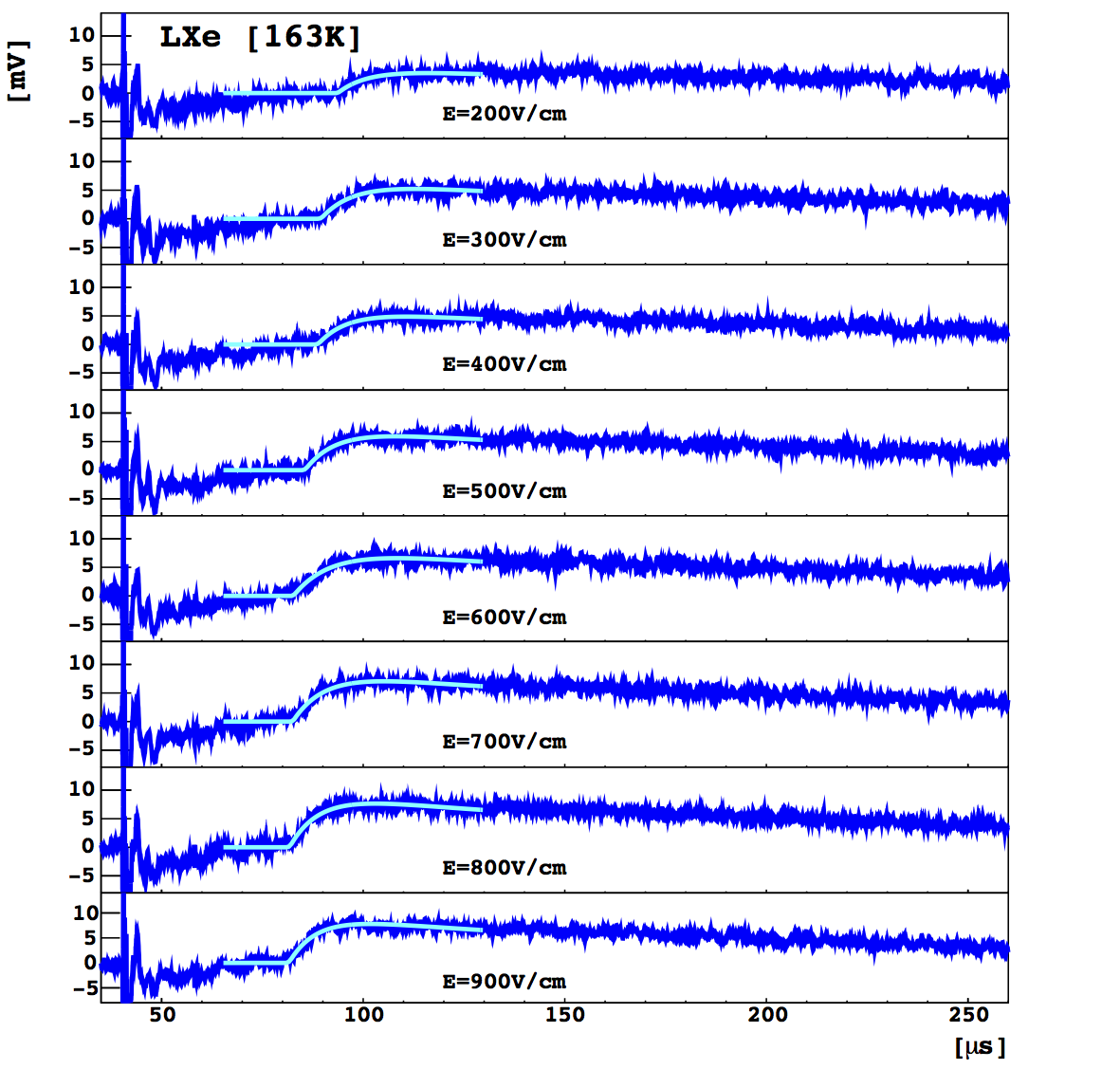}
\includegraphics[width=2.9in]{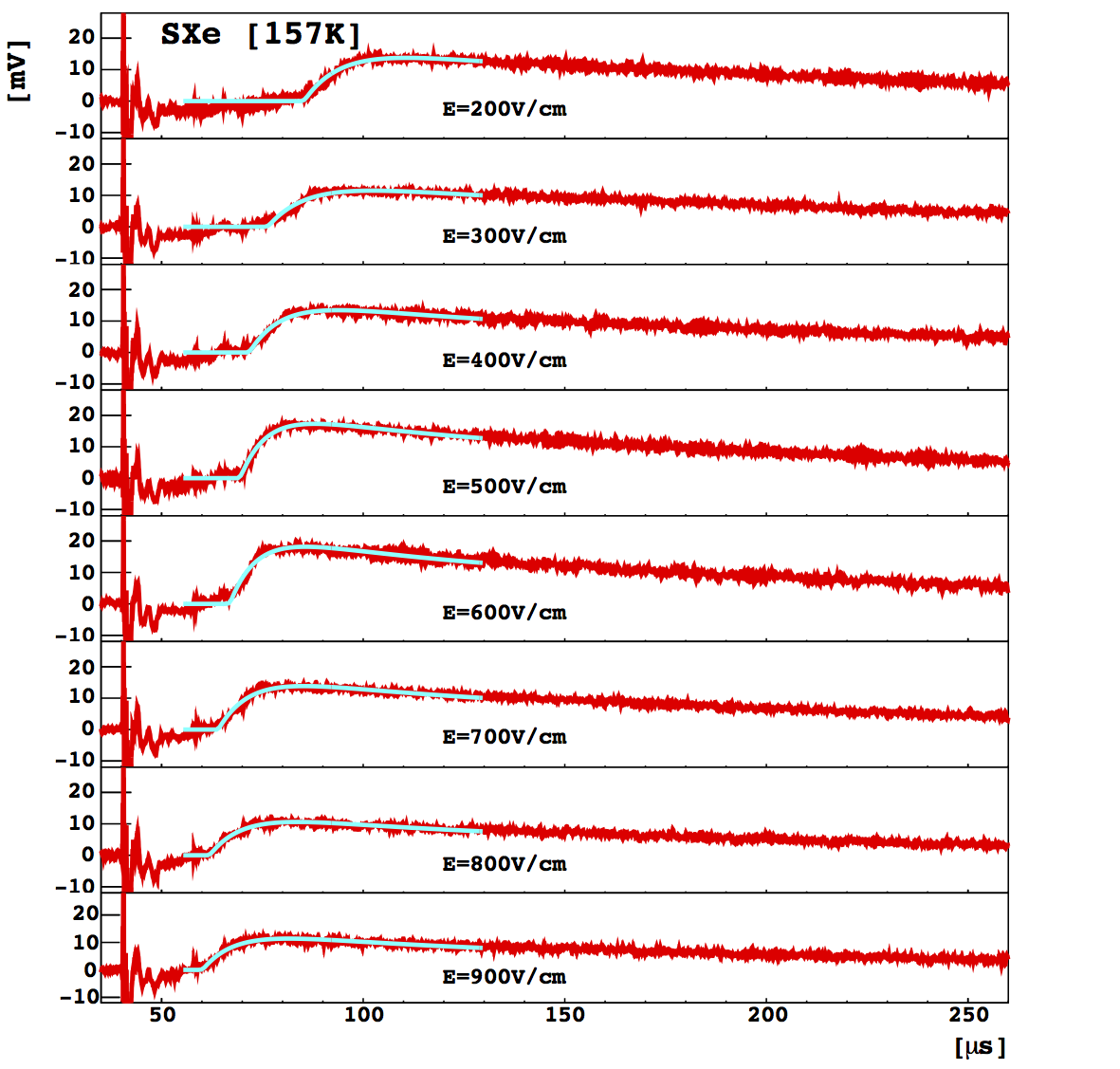}
\caption{Examples of pulse shapes at different drift fields for liquid xenon (left panel) and for solid xenon (right panel). The fit results are overlaid on the traces in cyan curves.~\label{fig:tracesummary}}
\end{center}
\end{figure}

\par FIG.~\ref{fig:tracecomp} shows examples of measured traces in liquid (163\,K) and solid phase (157\,K) of xenon. The applied electric field is 500\,V/cm. The glitch peaks at 40\,$\mu$s are induced by the high voltage trigger of the pulsed light source unit, and they can be used as references of trigger time. A total of 64 traces are averaged to cancel white noise. Remaining noises are from the heater power source which periodically switches from AC to DC, and noise pickup from the high-speed vacuum turbo. There is a slight monotonic increase of charge signals immediately after the glitch peaks that we could not completely remove. We suspect there could be a small leak of electric fields between the two grids. It is clearly visible that the start time of the anode signal in the solid xenon (red curve) is earlier than that in the liquid xenon (blue curve). Therefore, a faster electron drift occurs in solid xenon compared to that in liquid xenon over a large distance (8.0\,cm) at the same 500\,V/cm electric field. To find the anode start time ($t_i$), a functional form fit is carried out to the anode trace, given as $f(t;A,t_i,\tau) = A/t \cdot (1-\exp(-(t-t_i)/\tau))$, when $t>t_i$ and $f(t)=0$ when $t\le t_i$ where $A$, $\tau$ and $t_i$ are free parameters. The baseline of the anode trace is normalized by averaging $t=10-30$\,$\mu$s of the pre-trigger region. The trigger signal as the electron creation is at $t_i=40 \pm 0.5 \mu$s. The electron drift time is defined by $t_d = t_0-t_i$. The drift velocity is defined by $v$=8.0\,cm/$t_d$. The signal amplitude at the anode in solid phase is 2.8 times larger than that in the liquid phase while the cathode signal amplitudes in both phase are similar. This observed difference of the anode amplitude can be attributed to the electron life time in the xenon. Assuming the amplitude of the trace is proportional to the collected charge~\cite{Adamowski:2014daa}, the electron life time ($\tau$) can be estimated by $\tau = t_d / \log(Q_c/Q_a)$, where $Q_a/Q_c \simeq V_a/V_c$. The $V_a$ ($V_c$) is the maximum (minimum) voltage of anode (cathode) trace, and $Q_a$ ($Q_c$) is the collected charge at the anode (cathode). The electron life time in liquid xenon is $\tau_L = 44.1 \mu s / \log (78.0\mbox{mV}/5.9\mbox{mV}) = 39.1 \mu s$, while that in solid is $\tau_S = 29.4 \mu s / \log (78.0\mbox{mV}/16.6\mbox{mV}) = 43.8 \mu s$. Therefore, the electron life time in solid and liquid phases is differ by only 12\%. As the anode signals are used to evaluate the drift time, the absolute value of the cathode and anode amplitudes do not seriously affect the electron drift time measurement.

\begin{table}[t!]
\begin{center}
\begin{tabular}{c | c | c | c}
\hline 
 E [V/cm] & ~~~~~$v_L$ [cm/$\mu$s]~~~~~ &~~~~~ $v_S$ [cm/$\mu$s]~~~~~&~~$v_S/v_L$~~\\
\hline
200 & 0.148 $\pm$ 0.003 & 0.181 $\pm$ 0.003 & 1.22 \\
250 & 0.157 $\pm$ 0.004 & 0.200 $\pm$ 0.004 & 1.27 \\
300 & 0.162 $\pm$ 0.003 & 0.227 $\pm$ 0.004 & 1.40 \\
350 & 0.166 $\pm$ 0.003 & 0.245 $\pm$ 0.005 & 1.48 \\
400 & 0.167 $\pm$ 0.003 & 0.255 $\pm$ 0.004 & 1.53 \\
450 & 0.174 $\pm$ 0.004 & 0.266 $\pm$ 0.004 & 1.53 \\
500 & 0.177 $\pm$ 0.003 & 0.274 $\pm$ 0.004 & 1.55 \\
550 & 0.180 $\pm$ 0.004 & 0.288 $\pm$ 0.005 & 1.60 \\
600 & 0.184 $\pm$ 0.003 & 0.302 $\pm$ 0.006 & 1.64 \\
650 & 0.186 $\pm$ 0.003 & 0.325 $\pm$ 0.006 & 1.74 \\
\hline
700 & 0.189 $\pm$ 0.006 & 0.343 $\pm$ 0.005 & 1.82 \\
750 & 0.191 $\pm$ 0.003 & 0.357 $\pm$ 0.005 & 1.87 \\ 
800 & 0.193 $\pm$ 0.003 & 0.382 $\pm$ 0.007 & 1.98 \\
850 & 0.194 $\pm$ 0.003 & 0.390 $\pm$ 0.007 & 2.01 \\ 
900 & 0.193 $\pm$ 0.003 & 0.397 $\pm$ 0.006 & 2.05 \\
\hline  
\end{tabular}
\caption{Mean values of the measured electron drift velocities for various electric fields in liquid ($v_L$) and solid ($v_S$) xenon. The errors are from the functional form fit of $t_0$ through multiple measurements. The anode-grid to anode voltages are set to 400\,V/cm at above 700\,V/cm of drift fields. See text for details.}\label{tab:drifttime}
\end{center}
\end{table}

\par FIG.~\ref{fig:tracesummary} shows examples of the anode signals in liquid (left) and solid (right) xenon at various applied electric fields, where the cyan curves are the fit functions. TABLE~\ref{tab:drifttime} summarizes these measurements. We used the fit function $f(t)$ in order to determine $t_i$. The fit range of the pulses in liquid phase is $t=65-130$\,$\mu$s and that of solid phase pulses is $t=55-130$\,$\mu$s. The fit results of $t_i$ are relatively insensitive to the choice of fit range as long as it is kept to $45<t<200$\,$\mu$s, which assures that the HV glitch and the pulse tail does not dominate the fitting results. The pre-anode signal region shows undershooting traces after the HV glitch and slowly approach the baseline at around 60$\mu$s. In order to understand systematic uncertainties associated with the pre-anode signal region to the $t_i$ measurements, we fit the pre-pulse region (50$\mu$s to $t_i$) using a second-order polynomial function and corrected the baseline of the trace, then repeated the $f(t)$ functional form fit process. The largest shift of the $t_i$ value is -0.27$\mu$s which occurs in the solid phase with 900\,V/cm, or +0.005\,cm/$\mu$s difference in the drift velocity. The uncertainty of the electron drift time is estimated through repeated measurements in the same electric field. The standard deviation from the repeated $t_i$ measurements is estimated about 2.5\%. The drift fields from the cathode to anode are set uniform up to 650\,V/cm. However, owing to the charge breakdown at the high voltage feedthrough above $\sim$5,000\,V, the electric field between anode-grid and anode voltages are set to 400\,V/cm above 700\,V/cm of drift fields configuration. Therefore, amplitudes of the anode pulses may be smaller than expected at those electric fields. We believe the $t_d$ measurements should not be seriously affected by these inconsistent electric fields. However, a decisive test will be possible after resolving the charge breakdown issue in future studies. After the first set of measurements (liquid and solid), we melted the solid xenon to liquid. The heating of the xenon chamber and convection of xenon increased the impurity level in the xenon bulk. Therefore, after the melting, the initial liquid xenon does not show large enough cathode and anode signal amplitudes. The liquid xenon was continuously purified until we recover the cathode and anode signal amplitudes of the first set of measurements. This re-purifying process took a few days to reach the saturation level of charge amplitude. Then the second set of measurements are then repeated in liquid and solid phases.

\begin{figure}[t!]
\begin{center}
\includegraphics[width=4.8in]{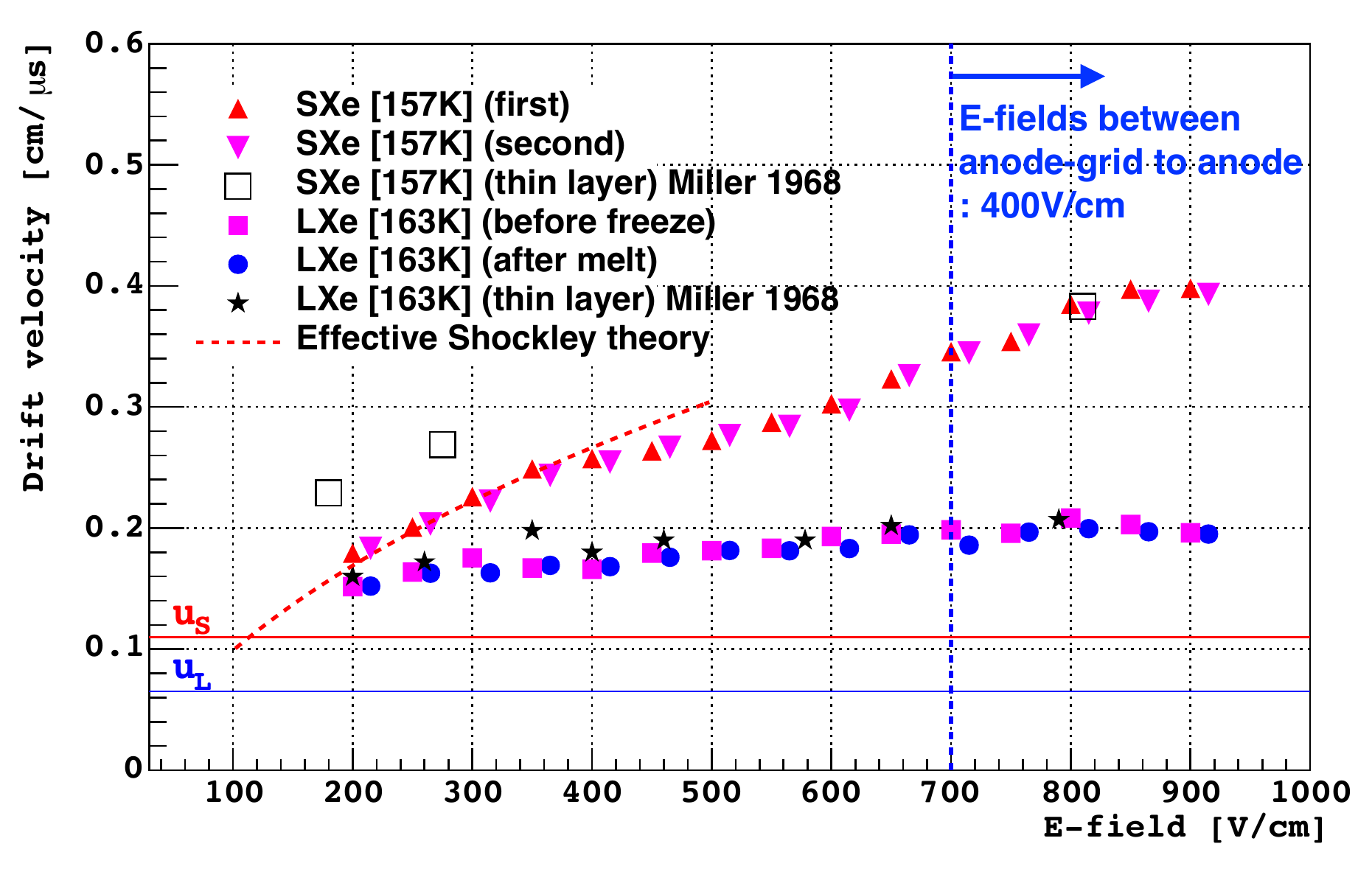}
\caption{Electron drift velocity in liquid and solid xenon for various electric field strength. The measured electron drift velocities in liquid xenon before the solidification are in magenta squares ({\textcolor{magenta}{\tiny$\blacksquare$}}). The measurements in solid phase for the first trial of solidification are in red triangles ({\textcolor{red}{\tiny\ding{115}}}). The measurements in liquid phase after melting the xenon are in blue circles ({\textcolor{blue}{\tiny\ding{108}}}), where a slight offset for the electric fields is just for display purpose. The second trial of solid xenon measurements after refreezing the xenon are in magenta down-triangles ({\textcolor{magenta}{\tiny\ding{116}}}), where the electric fields are off-set just for display purpose.. The previous measurements by Miller {\it et al.} in thin-layers (145$\mu$m to 228$\mu$m) of liquid and solid xenon are shown in stars ({\ding{72}}) and blank squares ($\square$) respectively~\cite{Miller:1968zza}. The red line ($u_S$) indicates the sound velocity in solid xenon (0.110\,cm/$\mu$s) at 157\,K. The blue line ($u_L$) indicates the sound velocity in liquid xenon (0.065\,cm/$\mu$s) at 163\,K. The red dashed curve shows the effective Shockley theory. See text for details.\label{fig:evelocity}}
\end{center}
\end{figure}

The measured drift velocity as a function of the electric field is shown in FIG.~\ref{fig:evelocity}. The electron drift velocities in the liquid phase are measured before and after the solidification in various electric field strengths. The ``after melt" electric fields are offset just for the display purpose. It can be seen that the two measurements in liquid xenon are in good agreement. The results are also compared with previous measurements in thin-layers (145$\mu$m to 228$\mu$m) of liquid xenon by Miller {\it et al.}~\cite{Miller:1968zza}. The data points of the thin-layer measurements are taken by reading FIG.~6 of reference~\cite{Miller:1968zza}. The electron drift velocities in the liquid phase are consistent between the thin-layers of xenon and the large scale xenon. The drift velocities of the first and the second solidifications of the xenon are in a very good agreement. The applied electric fields for the "second" measurements are the same with the "first" measurement, and are offset for display purpose. In the solid phase, the measured electron drift velocity is consistent with that of the thin-layer measurement at a higher electric field (800\,V/cm). The electron drift velocity in solid xenon is a factor of 2 larger than in liquid at or above E=800\,V/cm. However, there are noticeable discrepancies in the measured drift velocities at lower electric fields. The electron drift velocities are about 30\%  faster in thin-layer compared to that in the large scale solid xenon at the drift fields of 200$\sim$300\,V/cm.

\section{Discussions}\label{sec:discussions}
\par The dynamics of electron propagation in a pure single crystalline xenon is simply determined by the interaction with acoustic phonons. For condensed xenon, the possible occurrence of molecular impurities must be taken into account when determining the electron drift as its presence can lead to inelastic scattering of electrons that changes the energy distribution of the electrons. Impurities that are electro-negative would not change the energy distribution of the electrons but lead to a reduction of the total number of drifted electrons. In order to avoid contamination from the outgassing of the detector components, the xenon glass chamber with the electrode had been baked at 40\,C for two days and then the chamber was evacuated to below 10$^{-6}$\,Torr using a turbo system for two weeks. We used research grade xenon with 99.999\% of purity. The other components reported by the gas provider are: krypton ($<$ 1 ppm), water ($<$ 1 ppm), hydrogen ($<$ 0.5 ppm), oxygen ($<$ 1 ppm), nitrogen ($<$ 2 ppm), hydrocarbons ($<$ 1 ppm), tetrafluoromethane ($<$ 1 ppm), and carbon dioxide ($<$ 0.5 ppm). These contaminations are all below the capable range of our Residual Gas Analyzer (RGA) ($<$ 100 ppm). No other contamination above the RGA background level has been observed in the xenon gas. We regard the impurity level of the molecular components in our condensed xenon to be low enough that they do not significantly affect the dynamics of electrons.

\par An extensive systematic study of electron drift velocities in 1\,cm thick xenon substrates for various electric fields and temperatures has been carried out~\cite{Gushchin1982}. Their measured electron drift velocities in liquid xenon is about 15\% faster and about 55\% (at 200\,V/cm) to 23\% (at 900\,V/cm) faster in solid xenon compared to our results. The method applied in reference~\cite{Gushchin1982} is a so-called {\it modified Townsend pulse discharge method}, where a short pulse of x-rays are irradiated to an electrode and electrons produced in that electrode drift to the other electrode. The electron drift velocity is essentially estimated by measuring the rise time of the voltage pulse with various corrections, such as effective penetration depth of x-ray beam into the xenon, mean electron path before capture, charge amplifier time constant, extrapolation of the linear part of pulse rise time to non-linear regime. These corrections significantly affect the value of the electron drift velocity. On the other hand, our method directly measures the drift time of electrons in a distance of 8.0\,cm of by shielding the electric fields using grids. Therefore, our method does not rely on any significant corrections. 

\par The element xenon forms a relatively simple liquid as the binding forces between the atoms are symmetric van der Waals force and the local atomic order is determined by the close packing of ideal spheres. However, the atoms still move freely in the liquid phase with random motion while they form a face-centered cubic structure in the solid phase and freeze. The thin-layer of solid xenon substrates that were produced in the cryostat of reference~\cite{Miller:1968zza} would most probably have been a single crystal or was consisted of a few polycrystalline layers. If so, the electrons drifting in thin-layer of crystal would not suffer too much scattering at each crystal boundary. For a large scale solid xenon, one can produce clear, transparent solid xenon specimens which are virtually perfect crystals. Nevertheless, in most cases, they are polycrystalline and contain a large number of microscopic defects~\cite{RGSv2}. The drifting electrons would scatter and slow down at each polycrystalline grain boundaries and the defects. In weak drift fields the electron scattering at the grain boundaries would be significant as the probabilities of scattering of electrons would be high. However at strong drift fields where the electrons acquire significant energy above the potential barrier, the probabilities of electrons trans-passing the boundaries would improve~\cite{Warkusz1987}. Hence the observed electron drift velocity in polycrystalline would be close to that in a single crystalline. For example, reference~\cite{Horowitz2001185} demonstrated that the electron mobility is proportional to the grain size in a polycrystalline film transistor. It was also shown that the electron mobility in a single crystalline is about an order of magnitude larger than of that in polycrystalline. The overall electron mobility ($\mu$) can be estimated by taking a harmonic mean of electron mobility in the grain ($\mu_g$) and the boundary ($\mu_b$); $1/\mu=1/\mu_g+1/\mu_b$. Therefore, in cases where $\mu_g \gg \mu_b$, $\mu$ only depends on the charge transport across boundaries. As the drift field increases, the kinetic energy of electron overcomes the barrier energy and $\mu_b$ becomes comparable to $\mu_g$. The microscopic mechanism which is qualitatively described above would require further systematic studies using well defined crystalline xenon substrates. 

\par We assume the hot-electron theory of Shockley~\cite{Shockley1951} effectively account for the results in low electric fields~\footnote{The Shockley theory is limited to a smaller range of electric field and cannot explain the drift velocity saturation. A more advanced hot-electron theory by Cohen and Lekner~\cite{1967PhRv..158..305C} describes the electron drift velocity in liquid argon remarkably well over large electric field ranges up to 10,000\,V/cm, but this theory cannot adequately describe other noble elements. The agreements between the two theories are good in low electric fields. Therefore, we used the Shockley theory to estimate the effective electron mobility in the solid phase measurements. In the case of liquid phase measurements, however, a saturation of the electron drift velocity was observed in the lowest electric field. Therefore the Shockley theory is not applicable in liquid phase measurement.}. The theory describes the energy distribution of the hot-electrons by an effective temperature ($T_e$) and the lattice temperature ($T$). The ratio of $T_e/T=\gamma$ is related to the applied electric field as $\gamma^2 - \gamma = (3\pi/32)\cdot(\mu_0 E/u)^2$, where the $\mu_0$ is electron mobility and $u$ is the sound velocity in the substrate. The drift velocity is given by $v=\gamma^{-1/2}\mu_0 E$. At low drift fields, $\gamma\simeq 1$ and $v\simeq\mu_0E$. In high electric fields the drift velocity becomes proportional to $E^{1/2}$. In reference~\cite{Miller:1968zza}, the electron mobility ($\mu_0$) for liquid xenon is measured to be 2200\,cm$^2$/s/V at electric fields below 100\,V/cm. The saturation of the drift-velocity occurs at 4.4 times the sound velocity in liquid xenon ($u_L$) above 10\,kV/cm of the high-field region. The thin-layer of solid xenon study  sets the lower limit of the electron mobility at 3000\,cm$^2$/s/V by effectively fitting the Shockley theory curve to the measured drift velocities at low drift fields. The complete saturation of the drift-velocity occurs at 5.0 times the sound velocity in solid xenon ($u_S$) above 10\,kV/cm. Therefore, assuming the Shockley theory is effectively extended to the drift fields in the range of 200\,V/cm to 400\,V/cm, the electron mobility is estimated at 1100\,cm$^2$/s/V in the non-linear regime. The red-dashed curve in the FIG.~\ref{fig:evelocity} shows the fit results of the Shockley theory to the measured electron drift velocity in solid xenon.

\section{Conclusion}\label{sec:conclusion}
\par A study of electron drift in a large scale optically transparent solid xenon is reported. In the liquid phase (163\,K), the drift speed is 0.193 $\pm$ 0.003 cm/$\mu$s, while the drift speed in solid phase (157\,K) is 0.397 $\pm$ 0.006 cm/$\mu$s at an electric field strength of 900 V/cm. Therefore it has been demonstrated that the electron drift speed is faster by factor two in the solid phase compared to that in the liquid phase for large scale xenon. The results are consistent with the thin-layer measurements in the previous study ~\cite{Miller:1968zza}. However, in lower electric fields, the measured electron drift velocities are slower than that in the thin-layer substrate. Assuming the Shockley theory is effective in the non-linear regime, the electron mobility is estimated to be about 1100\,cm$^2$/s/V at 200\,V/cm to 400\,V/cm of electric fields in a large scale solid xenon.

\vspace{0.5cm}
\acknowledgments
\par We are very grateful to M.~Miyajima, J.~White, and A.~Bolozdnya for the initial discussions of the solid xenon particle detector and sharing their ideas. We would like to thank the Fermilab technical staff who aided in the design and construction of the apparatus. This work supported by the Department Of Energy Advanced Detector R\&D funding.

\bibliographystyle{h-physrev3}
\bibliography{sxref}{}
\end{document}